\documentclass[conference]{IEEEtran}
\ifCLASSINFOpdf
\else
\fi
\usepackage{graphicx}

\begin{document}
%
\title{Performance Analysis and Optimization of a Hybrid Distributed Reverse Time Migration Application}

\author{\IEEEauthorblockN{Sri Raj Paul, John Mellor-Crummey (Advisor)}
\IEEEauthorblockA{Rice University \\ Houston, Texas, USA\\
\{sriraj, johnmc\}@rice.edu}
\and
\IEEEauthorblockN{Mauricio Araya-Polo (Advisor), Detlef Hohl (Advisor)}
\IEEEauthorblockA{Shell international Exploration \& Production Inc. \\ Houston, Texas, USA\\
\{mauricio.araya, detlef.hohl\}@shell.com}}


%


\maketitle

\begin{abstract}
Applications to process seismic data employ scalable parallel systems to produce timely results. To fully exploit emerging processor architectures, application will need to employ threaded parallelism within a node and message passing across nodes. Today, MPI+OpenMP is the preferred programming model for this task. However, tuning hybrid programs for clusters is difficult. Performance tools can help users identify bottlenecks and uncover opportunities for improvement. This poster describes our experiences of applying Rice University's HPCToolkit and hardware performance counters to gain insight into an MPI+OpenMP code that performs Reverse Time Migration (RTM) on a cluster of multicore processors. The tools provided us with insights into the effectiveness of the domain decomposition strategy, the use of threaded parallelism, and functional unit utilization in individual cores. By applying insights obtained from the tools, we were able to improve the performance of the RTM code by roughly 30 percent.
\end{abstract}


%
\IEEEpeerreviewmaketitle

\section{Introduction}

Reverse Time Migration (RTM)~\cite{rtm} helps to create images of subsurface structures by simulating the propagation of pressure waves through them. RTM yields more accurate results than one-way Wave Equation Migration (WEM), but is computationally expensive (at least one order of magnitude more than WEM)~\cite{araya-polo}. One way to accelerate RTM is to distribute work among multiple nodes. Here, we describe our experiences analyzing and tuning a distributed RTM (DRTM) application developed at Shell Exploration \& Production Inc.
The code uses MPI message passing interface~\cite{mpi} for distributed-memory parallelism across nodes and OpenMP~\cite{openmp} for thread-level parallelism within nodes.

Tuning MPI+OpenMP programs for high performance is challenging. Understanding such applications requires analysis at multiple levels: 

\begin{itemize}
    \item domain decomposition and interprocess communication for a distributed-memory parallelization,
    \item threaded parallelism on a node, and
    \item functional unit and cache utilization within a core.
\end{itemize}

We used Rice University's HPCToolkit \cite{hpctoolkit} and hardware performance counters to gain insight into the DRTM application. HPCToolkit measures application performance using sampling to gather call stack profiles and call stack traces.
Based on insights gained, we were able to make improvements on each front.

\section{Structure of the Application}
\begin{figure}[htb]
    \centering
    \includegraphics[width=0.47\textwidth]{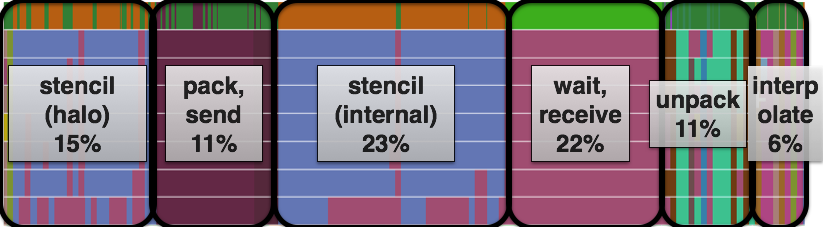}
    \caption{HPCToolkit showing the division of a single time step.}
    \label{fig:ht_timestep}
\end{figure}

The DRTM application is a PDE solver that applies a high order stencil to a 3D block of data in each time step. Each time step is divided into multiple phases as shown in Fig \ref{fig:ht_timestep} \footnote{A complete execution includes other activities such as reading data from input files and hence reported percentages do not add to 100\%.}.

The first phase of a time step performs a stencil computation on data in the halo regions. Then data points in halo regions are packed into messages that are later exchanged between neighbors. The DRTM code then initiates nonblocking MPI calls for point-to-point communication with the aim of overlapping communication with computation. Next, the code performs stencil computation for non-halo regions while communications are pending for data in the halo regions. After the stencil computation finishes, each MPI process waits for the arrival of halo data sent prior to the stencil computation. Finally the code unpacks data from the halo exchanges and performs interpolation where necessary.

\section{Performance Evaluation and Optimization}

Our initial measurements of the DRTM code with HPCToolkit showed that the code spends 7\% of its execution time on \texttt{memcpy} operations. Our investigation of this overhead led us to examine the abstraction layer that DRTM employs to support parallelization using different programming models, including OpenMP and CUDA \cite{cuda}. To accommodate accelerator programming models, which currently require copying data into a different memory space, the abstraction layer copied data even when using OpenMP, which does not require a copy because it executes in the same memory space. Refactoring the abstraction layer to copy pointers instead of employing a OpenMP loop nest to copy array data in parallel improved performance by roughly 10\%.

\begin{figure}[t]
    \centering
    \includegraphics[width=0.47\textwidth]{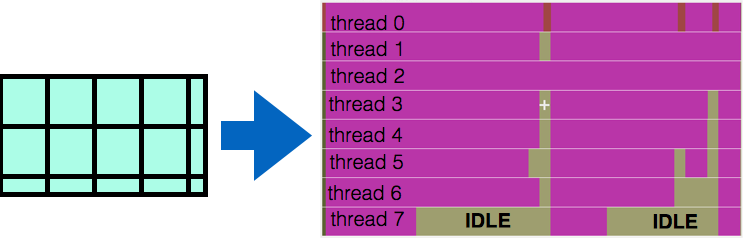}
    \caption{Smaller tiles at the high end of each dimension cause imbalance when OpenMP \texttt{static} scheduling is used for a collapsed 2D loop nest over the tiles.}
    \label{fig:static_partition}
\end{figure}

Examination of sample-based execution traces using HPCToolkit's \texttt{hpctraceviewer} \footnote{\texttt{hpctraceviewer} presents execution trace in a time-centric view.} revealed that some threads experience significant idleness during the RTM stencil calculation indicating an imbalance in work assigned to the threads. DRTM's stencil computation loop is tiled to improve cache reuse and a pair of loops over Y and Z tiles are collapsed into a single loop. Iterations of the loop over tiles are executed in parallel using a \texttt{static} schedule. In our investigation of the imbalance, we determined that the decomposition of work into tiles is as illustrated in Fig \ref{fig:static_partition}. 
OpenMP computes equal partitions of the iterations in the collapsed loop with no knowledge of tile sizes. A difference in tile sizes causes cores assigned a collection of small tiles to finish early. Changing the OpenMP scheduling strategy for this loop to \texttt{dynamic} reduced the imbalance and improved overall performance by roughly 5\%.

Even though MPI communication is performed with non-blocking sends initiated before and completed after the stencil computation, processes stall after the stencil computation waiting for messages to arrive. This wait caused us to question whether the communication was making progress during the stencil computation. Making use of the asynchronous progress engine provided by Intel's MPI library did not improve performance. To enable true overlap, we had a single application-level thread serve as a communication thread and perform all MPI communication~\cite{bamboo,asyncmpi} which helped to improve performance by around 10\%.

After adjusting the domain decomposition and loop parallelization strategies, we next checked how functional units are being utilized. We added code to measure various characteristics of the stencil code's execution using hardware performance counters. These measurements led us to discover that significant number of cache misses are occurring. Originally, the 3D stencil computation was performed by iterating over the 3D data in Y-Z-X order, where X is the innermost loop. The stencil computation for one point uses 49 points in the X-Y plane but only 41 points in the X-Z plane. Interchanging the Y and Z loops reduced L2 cache misses by increasing reuse and improved performance by roughly 5\%.

\begin{figure}[t]
    \centering
    \includegraphics[width=0.45\textwidth]{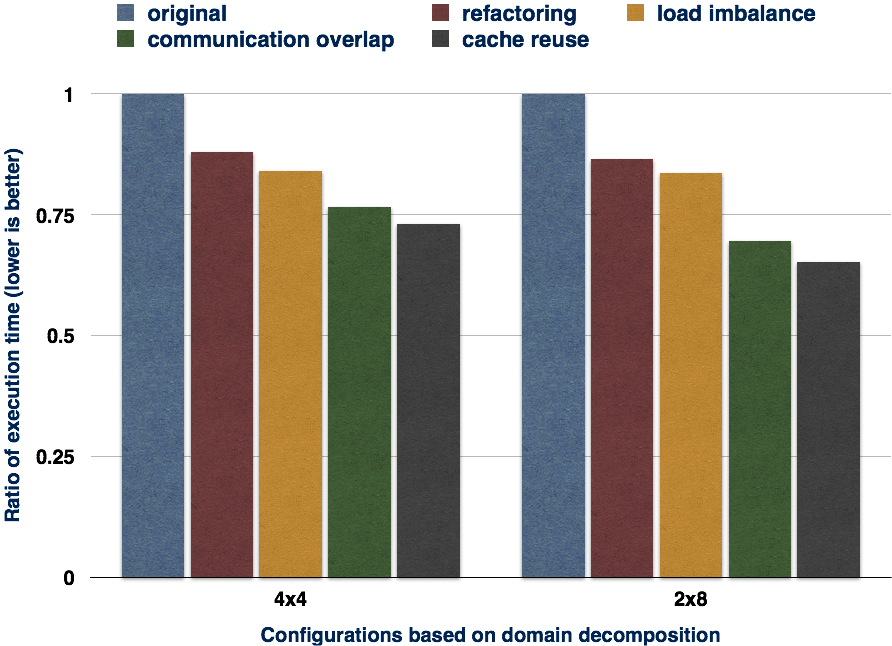}
    \caption{Improvement with respect to each optimization for two configurations: 4x4 and 2x8 domain decompositions of the X-Y plane.}
    \label{fig:performance}
\end{figure}

A bar graph showing performance improvement w.r.t to each optimization is shown in Fig \ref{fig:performance}.

\section{Conclusions}

Insights provided by HPCToolkit and hardware performance counters helped us to improve the performance of the DRTM code implemented using MPI+OpenMP significantly, improving load balance across MPI ranks, between threads, as well as improving the efficiency of computations on individual cores by avoiding unnecessary copies and reducing cache misses. Optimizations based on these insights improved the performance of DRTM by about 30\% on 16 MPI processes with eight threads per process running on 8 nodes of a cluster.

\section*{Acknowledgment}
This work was partially supported by Shell International Exploration \& Production Inc. under research agreement PT46021.



\bibliographystyle{IEEEtran}
\bibliography{IEEEabrv,sigproc}
%



\end{document}